\documentclass[iop]{emulateapj}
\usepackage{amsfonts,amsmath,amssymb,ulem,nicefrac,verbatim,color}
\usepackage{psfig,array,graphicx}

\def\simlt{\mathrel{\hbox{\rlap{\hbox{\lower4pt\hbox{$\sim$}}}\hbox{$<$}}}}
\def\simgt{\mathrel{\hbox{\rlap{\hbox{\lower4pt\hbox{$\sim$}}}\hbox{$>$}}}}

\def\ale{\mathrel{\hbox{\rlap{\hbox{\lower4pt\hbox{$\sim$}}}\hbox{$<$}}}}
\def\age{\mathrel{\hbox{\rlap{\hbox{\lower4pt\hbox{$\sim$}}}\hbox{$>$}}}}

\def\ra#1#2#3{#1$^{\rm h}$#2$^{\rm m}$#3$^{\rm s}$}
\def\dec#1#2#3{$#1^\circ#2'#3''$}

\def\spose#1{\hbox to 0pt{#1\hss}}

\begin{document}

\title{Discovery of an outflow from radio observations of the tidal disruption event ASASSN-14\lowercase{li}}

\author{K.~D.~Alexander, E.~Berger, J.~Guillochon, B.~A.~Zauderer, \& P.~K.~G.~Williams}
\affiliation{Harvard-Smithsonian Center for Astrophysics, 60 Garden St., Cambridge, MA 02138, USA}

\begin{abstract}
We report the discovery of transient radio emission from the nearby optically-discovered TDE ASASSN-14li (distance of 90 Mpc), making it the first typical TDE detected in the radio, and unambiguously pointing to the formation of a non-relativistic outflow with a kinetic energy of $\approx 4-10\times10^{47}$ erg, a velocity of $\approx 12,000-36,000$ km s$^{-1}$, and a mass of $\approx 3\times10^{-5}-7\times10^{-4}$ M$_{\odot}$.  We show that the outflow was ejected on 2014 August 11--25, in agreement with an independent estimate of the timing of super-Eddington accretion based on the optical, UV, and X-ray observations, and that the ejected mass corresponds to about $1-10\%$ of the mass accreted in the super-Eddington phase.  The temporal evolution of the radio emission also uncovers the circumnuclear density profile, $\rho(R)\propto R^{-2.5}$ on a scale of about 0.01 pc, a scale that cannot be probed via direct measurements even in the nearest SMBHs.  Our discovery of radio emission from the nearest well-studied TDE to date, with a radio luminosity lower than all previous limits, indicates that non-relativistic outflows are ubiquitous in TDEs, and that future, more sensitive, radio surveys will uncover similar events.
\smallskip
\end{abstract}

\keywords{accretion, accretion disks --- black hole physics --- galaxies: nuclei --- radiation mechanisms: non-thermal --- radio continuum: galaxies --- relativistic processes}

\section{Introduction}
The tidal disruption of stars by supermassive black holes (SMBH) lights up dormant systems and can be used to probe accretion and outflow processes.  Theoretical calculations indicate that most tidal disruption events (TDEs) lead to super-Eddington fallback, which in turn drives outflows \citep{rees88,ek89,sq09,gr13}.  The discovery of luminous radio emission from the $\gamma$-ray TDE Sw\,J1644+57 revealed the formation of a relativistic jetted outflow \citep{zbs+11,bzp+12}, but such events represent at most a few percent of the TDE population \citep{zbs+11,bgm+11,bkg+11,mgm+15}. While the sample of well-studied TDE candidates has expanded greatly in recent years, direct evidence for outflows in the bulk of the TDE population, discovered through optical, ultraviolet (UV), and X-ray observations, has been lacking.  

Radio observations are an ideal way to search for outflows in TDEs, as radio emission is expected to persist for months or years after the event even if the jet's orientation is off-axis. Most TDEs detected within the past decade have been followed up in the radio, but no ``typical" TDEs (i.e. those lacking $\gamma$-ray and hard X-ray emission) have been convincingly detected \citep{bmc+13,vfk+13}. (Weak radio emission has been seen in one or two TDE host galaxies, but the emission does not appear to be transient and these detections have been attributed to AGN activity; \citealt{vfk+13}.) Furthermore, due to the large distances of most TDEs discovered to date, the resulting upper limits are only able to rule out the presence of off-axis relativistic jets similar to those observed in gamma ray bursts or in Sw\,J1644+57 \citep{vfk+13,cbg+14}. The existence of lower energy, non-relativistic outflows cannot be ruled out by these observations.

On 2014 November 22, the All Sky Automated Survey for SuperNovae (ASAS-SN) reported the discovery of the new transient ASASSN-14li, coincident with the nucleus of the nearby galaxy PGC\,043234 (redshift $z = 0.0206$ luminosity distance $d_L\approx90$ Mpc).  Extensive optical, UV, and X-ray follow-up have confirmed that ASASSN-14li can be consistently modeled as a TDE, and is atypical for an AGN flare or supernova \citep{hol15,mkm15}. In this paper, we report the discovery and follow-up of transient radio emission from ASASSN-14li. The transient nature of the radio emission was independently reported by \cite{vv15}, although most of their observations were taken at a single frequency, strongly limiting their ability to constrain the evolution of the spectral energy distribution (SED). 

\begin{figure*} 
\centerline{\includegraphics[width=\textwidth]{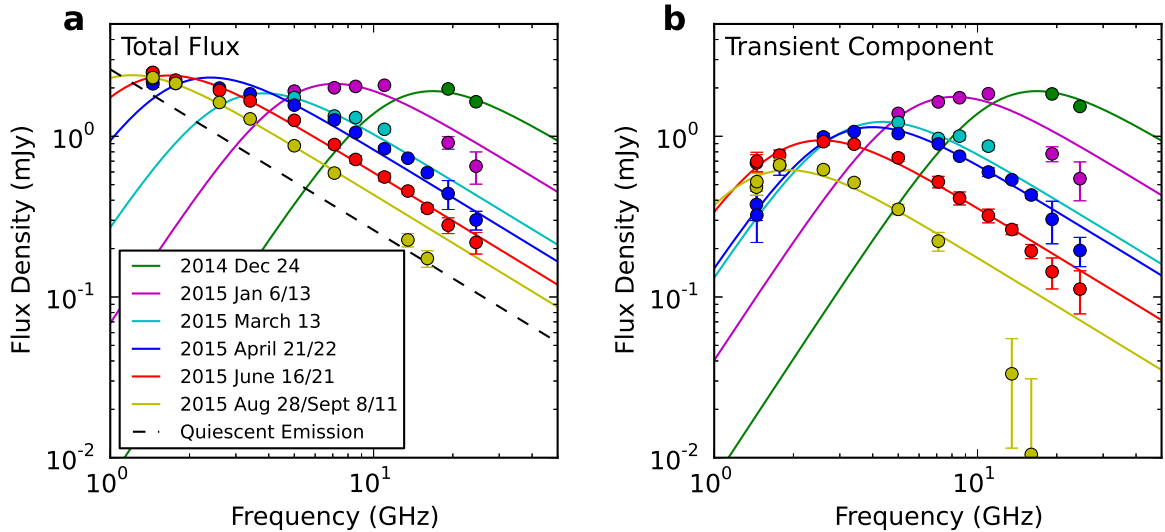}} 
\caption{Radio observations of the TDE ASASSN-14li spanning December 2014 to September 2015. Filled circles mark the observed radio flux densities (in many cases, the errorbars, which correspond to 1 standard deviation, are smaller than the points; Table 1), while solid lines are best-fit models for synchrotron emission from a power-law distribution of electrons \citep{gs02,dnp13}, $N(\gamma)\propto\gamma^{-3}$ for $\gamma\geq\gamma_m$ (Section \ref{sec:mod}). (a) The total flux observed at each frequency.  The dashed black line indicates a $F_\nu\propto \nu^{-1}$ power law model for the underlying quiescent emission component, whose existence is implied by the archival radio detections. (b) Residual transient radio flux density obtained by subtracting the modeled quiescent emission component.  These residual flux densities have a spectral shape characteristic of a synchrotron self-absorbed spectrum, with a spectral slope of $F_\nu\propto \nu^{5/2}$ below the peak and $F_\nu\propto \nu^{-1}$ above the peak.  The evolution of the SED is typical of synchrotron emission from an expanding outflow.  We note that our 2014 December 24 observations only weakly constrain the location of the spectral peak, so all parameters inferred for this epoch are considered to be lower limits.}
\label{fig:sed}
\end{figure*}

The rest of this paper is structured as follows. In Section \ref{sec:obs}, we present our radio observations of ASASSN-14li. In Section \ref{sec:arc}, we discuss archival observations of ASASSN-14li's host galaxy PGC\,043234 to provide a context for our modeling. In Section \ref{sec:mod}, we outline our model for the radio emission and use it to infer physical properties of the outflow launched by the TDE and the pre-event circumnuclear density. In Section 5, we compare our results to independent modeling of the X-ray, UV, and optical observations of ASASSN-14li and address alternate explanations for the emission. We conclude in Section \ref{sec:conc}.

\section{Radio Observations and Data Analysis}\label{sec:obs}

Following the optical discovery of ASASSN-14li, we initiated radio follow-up observations with the Karl G. Jansky Very Large Array (VLA) on 2014 December 24 at a frequency of 21.8 GHz and detected a source with a flux density of $1.85\pm 0.03$ mJy.  The position of the radio source, $\alpha_{\rm J2000}=$\ra{12}{48}{15.226}, $\delta_{\rm J2000}=$\dec{+17}{46}{26.47} ($\pm 0.01$ arcsec), is consistent with the optical position. We continued to monitor the source and obtained six epochs of observations spaced at $1-2$ month intervals between 2014 December 24 and 2015 September 11 UT. Our observations span frequencies between 1.45 GHz and 24.5 GHz and reveal significant fading at high frequencies, a steady decline in the peak of the radio SED as a function of time (to $\approx 2$ GHz by September 2015), and a spectral slope of $F_\nu\propto\nu^{-1}$ above the peak frequency (Figure \ref{fig:sed}). These properties are typical of synchrotron emission from an expanding outflow.  

All radio observations were obtained with the VLA in the A, B, C, and intermediate configurations (program codes 14B-493 and 15A-476).  For all epochs and frequencies, we used 3C\,286 for bandpass and flux density calibration, and J1254+1141 for phase calibration.  We processed and imaged the data using the Common Astronomy Software Applications (CASA) software package \citep{mws+07}.  The flux densities and associated uncertainties were determined using the {\tt imtool} program within the {\tt pwkit} package\footnote{Available at {\tt https://github.com/pkgw/pwkit}} (version 0.6.99) and are summarized in Table~\ref{tab:obs}. The time evolution of the radio SED is also shown in Figure \ref{fig:sed}.

\section{Archival Radio Observations and Arguments Against an AGN Flare Origin for the Radio Emission from ASASSN-14li}\label{sec:arc}

The host galaxy of ASASSN-14li was previously detected in the NVSS (December 1993) and FIRST (November 1999) 1.4 GHz radio surveys \citep{bec95,con98}.  The FIRST and NVSS flux densities are $2.96\pm 0.15$ mJy and $3.2\pm0.4$ mJy respectively, corresponding to a radio luminosity of $L_\nu(1.4\,{\rm GHz})\approx 3\times 10^{28}$ erg s$^{-1}$ Hz$^{-1}$. If this radio emission is due to star formation activity in the host galaxy, then the inferred star formation rate is ${\rm SFR}\approx 2$ M$_{\odot}$ yr$^{-1}$ \citep{yc02}.  However, this is ruled out by archival optical, near-infrared, and far-infrared (FIR) observations of the host galaxy, which indicate that ${\rm SFR}\simlt 0.1$ M$_{\odot}$ yr$^{-1}$, and that the observed emission violates the radio-FIR correlation of star forming galaxies \citep{hol15}.  Thus, the radio emission is more likely due to a weak AGN, and indeed the archival radio luminosity places the host galaxy in the range of luminosities observed in low-luminosity Seyfert galaxies \citep{ho01}.

Our brightest 1.45 GHz flux density measurement constrains the maximum brightness of the quiescent component to be $\simlt 2$ mJy, indicating that the archival source has declined in brightness by about $30\%$ over the 16-year period between the FIRST measurement and our observations. This is typical of long-term AGN variability \citep{hov08}.  It is clear, however, that the event ASASSN-14li has more in common with previously-studied TDEs than with typical AGN flares.  Optical spectra and UV/optical imaging obtained during the outburst show strong blue continuum emission and broad hydrogen and helium emission lines, consistent with previously-observed TDEs and inconsistent with the evolution expected for an AGN or a supernova \citep{hol15}. Furthermore, the dramatic change in brightness we observe at our highest radio frequencies -- an order of magnitude decline over an 9 month period -- is much larger and more rapid than the radio variability observed in typical AGN flares, and is only comparable to the most extreme flares observed in BL Lacertae Objects \citep{hov08,niep09}.  Our radio spectral energy distributions of ASASSN-14li are also steeper in both the optically-thick ($F_{\nu} \propto \nu^{2.5}$) and optically-thin ($F_{\nu} \propto \nu^{-1}$) portions compared to typical AGN flares, which exhibit an average rising power law of $F_{\nu} \propto \nu^{0.4}$ and a declining power law of $F_{\nu} \propto \nu^{-0.2}$ \citep{hov08}.

\begin{center}
\setlength\LTcapwidth{2.5in}
\begin{longtable}{lccc}
\caption{Radio Observations}
\label{tab:obs} \\
\hline
\hline\noalign{\smallskip}
UT Date & $\Delta t$ & $\nu$ & $F_\nu$ \\
             & (days)         & (GHz)  & (mJy)   \\
\hline\noalign{\smallskip}
Dec 24.69 & 128.69 & 19.2 & 1.97 $\pm$ 0.03 \\ 
Dec 24.69 & 128.69 & 24.5 & 1.64 $\pm$ 0.03 \\ 
\hline\noalign{\smallskip}
Jan 6.38 & 141.38 & 5.0 & 1.91 $\pm$ 0.03 \\ 
Jan 6.38 & 141.38 & 7.1 & 2.00 $\pm$ 0.02 \\ 
Jan 6.38 & 141.38 & 8.5 & 2.04 $\pm$ 0.04 \\ 
Jan 6.38 & 141.38 & 11.0 & 2.08 $\pm$ 0.04 \\ 
Jan 13.32 & 148.32 & 19.2 & 0.91 $\pm$ 0.08 \\ 
Jan 13.32 & 148.32 & 24.5 & 0.65 $\pm$ 0.15 \\ 
\hline\noalign{\smallskip}
Mar 13.33 & 207.33 & 5.0 & 1.74 $\pm$ 0.02 \\ 
Mar 13.33 & 207.33 & 7.1 & 1.34 $\pm$ 0.02 \\ 
Mar 13.33 & 207.33 & 8.5 & 1.31 $\pm$ 0.06 \\ 
Mar 13.33 & 207.33 & 11.0 & 1.11 $\pm$ 0.05 \\ 
\hline\noalign{\smallskip}
Apr 21.25 & 246.25 & 1.4 & 2.18 $\pm$ 0.08 \\ 
Apr 21.25 & 246.25 & 1.5 & 2.12 $\pm$ 0.10 \\ 
Apr 21.25 & 246.25 & 1.8 & 2.13 $\pm$ 0.09 \\ 
Apr 21.25 & 246.25 & 2.6 & 2.00 $\pm$ 0.05 \\ 
Apr 21.25 & 246.25 & 3.4 & 1.84 $\pm$ 0.03 \\ 
Apr 21.25 & 246.25 & 5.0 & 1.56 $\pm$ 0.03 \\ 
Apr 21.25 & 246.25 & 7.1 & 1.26 $\pm$ 0.03 \\ 
Apr 22.21 & 247.21 & 8.5 & 1.06 $\pm$ 0.02 \\ 
Apr 22.21 & 247.21 & 11.0 & 0.84 $\pm$ 0.04 \\ 
Apr 22.21 & 247.21 & 13.5 & 0.73 $\pm$ 0.02 \\ 
Apr 22.21 & 247.21 & 16.0 & 0.59 $\pm$ 0.02 \\ 
Apr 22.21 & 247.21 & 19.2 & 0.44 $\pm$ 0.09 \\ 
Apr 22.21 & 247.21 & 24.5 & 0.30 $\pm$ 0.04 \\ 
\hline\noalign{\smallskip}
Jun 17.01 & 303.01 & 1.4 & 2.49 $\pm$ 0.09 \\ 
Jun 17.01 & 303.01 & 1.5 & 2.50 $\pm$ 0.10 \\ 
Jun 17.01 & 303.01 & 1.8 & 2.24 $\pm$ 0.06 \\ 
Jun 17.01 & 303.01 & 2.6 & 1.93 $\pm$ 0.04 \\ 
Jun 17.01 & 303.01 & 3.4 & 1.66 $\pm$ 0.04 \\ 
Jun 17.01 & 303.01 & 5.0 & 1.26 $\pm$ 0.04 \\ 
Jun 17.01 & 303.01 & 7.1 & 0.89 $\pm$ 0.04 \\ 
Jun 21.08 & 307.08 & 8.5 & 0.72 $\pm$ 0.04 \\ 
Jun 21.08 & 307.08 & 11.0 & 0.56 $\pm$ 0.03 \\ 
Jun 21.08 & 307.08 & 13.5 & 0.46 $\pm$ 0.02 \\ 
Jun 21.08 & 307.08 & 16.0 & 0.36 $\pm$ 0.02 \\ 
Jun 21.08 & 307.08 & 19.2 & 0.28 $\pm$ 0.03 \\ 
Jun 21.08 & 307.08 & 24.5 & 0.22 $\pm$ 0.03 \\ 
\hline\noalign{\smallskip}
Aug 28.94 & 375.94 & 1.4 & 2.15 $\pm$ 0.07 \\ 
Aug 28.94 & 375.94 & 1.5 & 2.22 $\pm$ 0.08 \\ 
Aug 28.94 & 375.94 & 1.8 & 2.13 $\pm$ 0.07 \\ 
Aug 28.94 & 375.94 & 2.6 & 1.58 $\pm$ 0.05 \\ 
Aug 28.94 & 375.94 & 3.4 & 1.26 $\pm$ 0.04 \\ 
Aug 28.94 & 375.94 & 5.0 & 0.81 $\pm$ 0.06 \\ 
Aug 28.94 & 375.94 & 7.1 & 0.49 $\pm$ 0.07 \\ 
Sep 8.96 & 386.96 & 1.4 & 2.49 $\pm$ 0.08 \\ 
Sep 8.96 & 386.96 & 1.5 & 2.49 $\pm$ 0.11 \\ 
Sep 8.96 & 386.96 & 1.8 & 2.15 $\pm$ 0.09 \\ 
Sep 8.96 & 386.96 & 2.6 & 1.65 $\pm$ 0.04 \\ 
Sep 8.96 & 386.96 & 3.4 & 1.30 $\pm$ 0.04 \\ 
Sep 8.96 & 386.96 & 5.0 & 0.89 $\pm$ 0.03 \\ 
Sep 8.96 & 386.96 & 7.1 & 0.61 $\pm$ 0.03 \\ 
Sep 11.92 & 389.92 & 13.5 & 0.23 $\pm$ 0.02 \\ 
Sep 11.92 & 389.92 & 16.0 & 0.17 $\pm$ 0.02 \\  
\hline\noalign{\smallskip}
\caption[]{Radio observations of ASASSN-14li.  All values of $\Delta t$ are relative to 2014 August 18.00 UT, the mean outflow launch date estimated from our modeling.}
\end{longtable}
\end{center}

\begin{figure*}
\centerline{\includegraphics[width=6.8in]{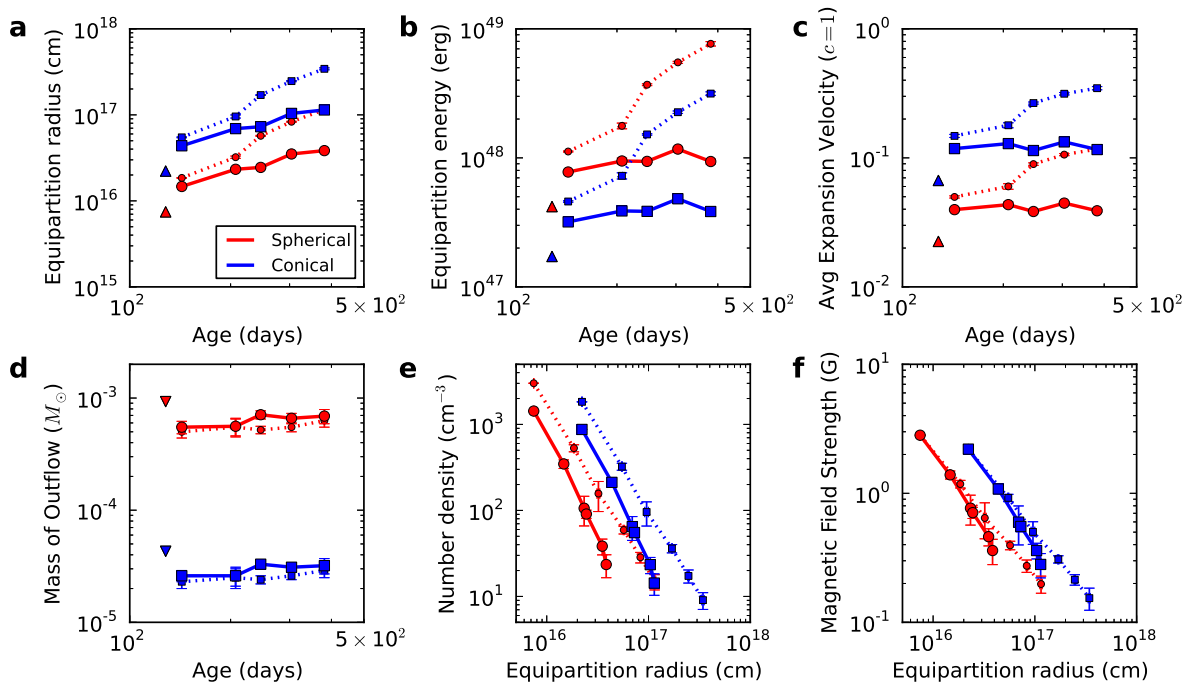}}
\caption{The temporal and radial dependencies of several physical quantities of the outflow inferred from synchrotron equipartition model fits to our radio observations. In each panel the dotted and solid lines mark the fits to the total radio flux densities (Figure 1, panel (a)) and transient flux density only (Figure 1, panel (b)), respectively.  The red circles mark the results for a spherical outflow while the blue squares mark the results for a conical outflow with a covering fraction of $10\%$.  We determine the radius of the emitting region as a function of time (a), the outflow kinetic energy as a function of time (b), the outflow expansion velocity as a function of time (c), the outflow mass as a function of time (d), the circumnuclear radial density profile (e), and the magnetic field radial profile (f).  The errorbars on the data points in each panel correspond to 1 standard deviation and are computed using a Markov Chain Monte Carlo approach that takes into account the uncertainties in the synchrotron model parameters.  The inferred quantities are summarized in Table 2.}
\label{fig:params}
\end{figure*}

Motivated by the archival radio detections, we assume that some portion of the radio emission we observe is due to a steady source not associated with the TDE. For simplicity, we assume that this component is constant in time for the period of our observations and follows a single power law shape, which we find to be $F_\nu\approx 1.8\,{\rm mJy}\,\,(\nu/1.4\,{\rm GHz})^{-1}$, accounting for about $80\%$ of our measured flux density at 1.4 GHz.  This spectral index is typical of at least some AGN of comparable luminosity in quiescence \citep{ho01}.  We subtract this model from our observed flux densities (Figure~\ref{fig:sed}(a)) and find that the remaining transient component exhibits a synchrotron self-absorbed spectral shape ($F_\nu\propto\nu^{5/2}$) below the peak frequency (Figure~\ref{fig:sed}(b)).  We model the SED of the transient source at each epoch of observations using the standard synchrotron equipartition model outlined in Section \ref{sec:mod} \citep{sr77,dnp13}.  For completeness, we also model the emission assuming that all of the flux we detect originates in a single component associated with the TDE, but find that this model provides a worse fit to the data, does not explain the archival radio detections, and leads to other inconsistencies (Section \ref{sec:1comp}); however, we note that the results of this model do not alter the basic conclusions of our analysis.

\section{Synchrotron Emission Model}\label{sec:mod}

We model our radio data with the standard synchrotron emission model, in which the blastwave generated by the outflow amplifies the magnetic field and accelerates the ambient electrons into a power law distribution, $N(\gamma) \propto \gamma^{-p}$ for $\gamma \geq \gamma_m$; here, $\gamma$ is the electron Lorentz factor, $\gamma_m$ is the minimum Lorentz factor of the distribution, and $p$ is the power law index.  This is the same model used to fit the radio emission from the relativistic TDE Sw\,J1644+57 \citep{zbs+11,bzp+12,zbm+13}, as well as from core-collapse SNe and GRBs.  We follow the procedures of \cite{dnp13} by assuming the outflow energy is minimized when the electron and magnetic field energy densities are in equipartition \citep{pac70,sr77,chev98}. Given the shape of the observed SEDs, we associate the peak frequency $\nu_p$ with the synchrotron self-absorption frequency $\nu_a$ and assume that the frequency corresponding to $\gamma_m$ is $\nu_m\simlt\nu_a$; this is generally the case for non-relativistic outflows \citep{dnp13}.  A comparison of the observed ($F_\nu\propto \nu^{-1}$) and model ($F_{\nu} \propto \nu^{-(p-1)/2}$) optically-thin power laws indicates that $p\approx 3$ \citep{gs02}. We further build on the results from modeling of radio emission in other transients to assume that the fraction of energy in the relativistic electrons \citep{dnp13} is $\epsilon_e=0.1$, and that the kinetic energy is dominated by protons.

The minimum energy analysis can also accommodate a non-spherical outflow, characterized by emitting area and volume fractions of $f_A \equiv A/\pi R^2$ and $f_V \equiv V/\pi R^3$, respectively; the spherical case corresponds to $f_A=1$ and $f_V=4/3$.  We explore two models, with $f_A=1$ (spherical outflow) and $f_A=0.1$ (conical outflow) to assess the effects of mild collimation, and we further assume that the emission emanates from a shell with a thickness of $0.1$ of the blastwave radius.

With this setup we can directly infer the equipartition radius $R_{\rm eq}$ and kinetic energy $E_{\rm eq}$ from the observed values of $\nu_p$ and $F_{\nu,p}$ at each epoch \citep{dnp13}:
\begin{eqnarray*}
R_{\rm eq}&=&(3.2\times10^{15}\,{\rm cm}) F_{\nu,p,mJy}^{\frac{9}{19}}d_{L,26}^{\frac{18}{19}} \nu_{p,10}^{-1} \\
&&\times \; (1+z)^{-\frac{10}{19}}f_A^{-\frac{8}{19}}f_V^{-\frac{1}{19}}\\
E_{\rm eq}&=&(1.9\times10^{46}\,{\rm erg}) F_{\nu,p,mJy}^{\frac{23}{19}}d_{L,26}^{\frac{46}{19}}\nu_{p,10}^{-1} \\
&& \times \; (1+z)^{-\frac{42}{19}}f_A^{-\frac{12}{19}}f_V^{\frac{8}{19}}
\end{eqnarray*}
where we have scaled $\nu_p$ in units of 10 GHz, $F_{\nu,p}$ in units of mJy, and the luminosity distance ($d_L$) in units of $10^{26}$ cm. For the spherical nonrelativistic case, these equations should be multiplied by factors of $4^{1/19}$ and $4^{11/19}$ due to additional geometric effects. With the inferred values of $R_{\rm eq}$ and $E_{\rm eq}$ we can furthermore derive other physical properties of the system, notably the ambient density ($n$), the magnetic field strength ($B$), the outflow velocity ($v_{\rm ej}$, or $\beta_{\rm ej}$ when scaled to $c$), and the outflow mass ($M_{\rm ej}$), as well as their time and radial dependencies.  We refer the reader to \cite{dnp13} for the exact formulae.  The resulting parameters for our two models ($f_A=1$ and $0.1$) are listed in Table~\ref{tab:params} and the results are shown in Figure \ref{fig:params}. We derive the uncertainties on $\nu_p$ and $F_p$ for each epoch via a Markov Chain Monte Carlo fitting technique. The uncertainties on the derived parameters are then computed using standard propagation of error.

Using our model fits to the individual epochs of observations we robustly measure the source size and kinetic energy as functions of time
.  We find that for an assumed spherical geometry, the radio observations require a non-relativistic outflow with a steady velocity of $v_{\rm ej}\approx 12,000$ km s$^{-1}$, freely expanding ($R_{\rm ej}\propto t$) from a radius of $\approx 1.5\times 10^{16}$ cm (January 2015) to $\approx 3.8\times 10^{16}$ cm (August/September 2015).  This velocity is larger than the width of the hydrogen and helium emission lines in the optical spectra of ASASSN-14li \citep{hol15}, indicating that these lines do not originate in the outflow.  Using the observed radius and extrapolating the observed constant expansion rate backwards we infer that the outflow was launched on 2014 August 11--25.  This date range is consistent with an independent estimate of the period of super-Eddington accretion derived from optical, UV, and X-ray observations of the TDE, which gives 2014 June 1--July 10 as the onset of super-Eddington accretion and 2014 September 1--September 15 as the time of peak accretion rate (with a level of about 2.5 times the Eddington rate); see Section \ref{sec:uvot}.  We therefore conclude that the outflow is linked to the super-Eddington accretion phase, rather than to the unbound tidal debris, which were launched much earlier at the time of disruption.  We note that assuming a conical outflow with $f_A=0.1$ instead of a spherical geometry increases the inferred radius and expansion velocity by about a factor of 3 (Figure~\ref{fig:params}), but the outflow launch date remains essentially unchanged.

\clearpage
\setlength\LTcapwidth{7in}
\begin{center}
\begin{longtable*}{cccccccccc}
\caption{Best-Fit Model Parameters} 
\label{tab:params} \\
\hline
\hline\noalign{\smallskip}
Model & $\Delta t$ & $\nu_p$ & $F_p$ & $R_{\rm eq}$     & $E_{\rm eq}$       & $\beta_{\rm ej}$ & $n$            & $M_{\rm ej}$ & $B$ \\
          & (days)        & (GHz)      & (mJy) & ($10^{16}$ cm) & ($10^{47}$) erg &                         & (cm$^{-3}$) & $10^{-4} M_{\odot}$ & (G) \\
\hline\noalign{\smallskip}
&128 & $\simlt$ 16.8 & $\simgt$ 1.91 & $\simgt$ 0.745 & $\simgt$ 4.2 & $\simgt$ 0.023 & $\simgt$ 1430 & $\simlt$ 9.3 & $\simlt$ 2.82 \\ 
&143 & 8.20 $\pm$ 0.10 & 1.76 $\pm$ 0.01 & 1.47 $\pm$ 0.02 & 7.8 $\pm$ 0.1 & 0.040 $\pm$ 0.001 & 350 $\pm$ 40 & 6.0 $\pm$ 0.7 & 1.39 $\pm$ 0.10 \\ 
Spherical &207 & 4.37 $\pm$ 0.20 & 1.23 $\pm$ 0.03 & 2.33 $\pm$ 0.10 & 9.5 $\pm$ 0.5 & 0.043 $\pm$ 0.002 & 110 $\pm$ 40 & 6.0 $\pm$ 1.0 & 0.77 $\pm$ 0.20 \\ 
($f_A=1$)&246 & 4.00 $\pm$ 0.06 & 1.14 $\pm$ 0.01 & 2.45 $\pm$ 0.04 & 9.4 $\pm$ 0.2 & 0.038 $\pm$ 0.001 & 90 $\pm$ 10 & 7.0 $\pm$ 0.6 & 0.71 $\pm$ 0.07 \\ 
&304 & 2.55 $\pm$ 0.06 & 0.94 $\pm$ 0.02 & 3.51 $\pm$ 0.08 & 11.7 $\pm$ 0.4 & 0.045 $\pm$ 0.001 & 38 $\pm$ 8 & 7.0 $\pm$ 0.7 & 0.46 $\pm$ 0.07 \\ 
&381 & 1.91 $\pm$ 0.07 & 0.62 $\pm$ 0.02 & 3.84 $\pm$ 0.10 & 9.4 $\pm$ 0.4 & 0.039 $\pm$ 0.001 & 24 $\pm$ 7 & 7.0 $\pm$ 1.0 & 0.36 $\pm$ 0.08 \\ 
\hline\noalign{\smallskip}
&128 & $\simlt$ 16.80 & $\simgt$ 1.91 & $\simgt$ 2.22 & $\simgt$ 1.7 & $\simgt$ 0.067 & $\simgt$ 874 & $\simlt$ 0.4 & $\simlt$ 2.2 \\ 
&143 & 8.20 $\pm$ 0.10 & 1.76 $\pm$ 0.01 & 4.37 $\pm$ 0.06 & 3.19 $\pm$ 0.05 & 0.118 $\pm$ 0.004 & 210 $\pm$ 20 & 0.26 $\pm$ 0.03 & 1.08 $\pm$ 0.09\\ 
Conical &207 & 4.37 $\pm$ 0.20 & 1.23 $\pm$ 0.03 & 6.9 $\pm$ 0.3 & 3.9 $\pm$ 0.2 & 0.129 $\pm$ 0.006 & 60 $\pm$ 20 & 0.26 $\pm$ 0.05 & 0.6 $\pm$ 0.2 \\ 
($f_A=0.1$)&246 & 4.00 $\pm$ 0.06 & 1.14 $\pm$ 0.01 & 7.3 $\pm$ 0.1 & 3.85 $\pm$ 0.07 & 0.114 $\pm$ 0.003 & 55 $\pm$ 7 & 0.33 $\pm$ 0.03 & 0.55 $\pm$ 0.05 \\ 
&304 & 2.55 $\pm$ 0.06 & 0.94 $\pm$ 0.02 & 10.0 $\pm$ 0.2 & 4.8 $\pm$ 0.1 & 0.133 $\pm$ 0.004 & 23 $\pm$ 5 & 0.31 $\pm$ 0.03 & 0.36 $\pm$ 0.05 \\ 
&381 & 1.91 $\pm$ 0.07 & 0.62 $\pm$ 0.02 & 11.4 $\pm$ 0.4 & 3.8 $\pm$ 0.2 & 0.116 $\pm$ 0.004 & 14 $\pm$ 4 & 0.32 $\pm$ 0.05 & 028 $\pm$ 0.06 \\ 
\hline\noalign{\smallskip}
\caption[]{Physical parameters of the outflow and circumnuclear environment derived from the synchrotron equipartition model that provides the best fit to our radio observations of ASASSN-14li.  We fit only the transient component of the radio fluxes.  We show values for two possible geometries: a spherical outflow ($f_A=1$) and a conical outflow with a covering fraction of $10\%$ ($f_A=0.1$). In both cases, we assume that the emitting region is a shell of thickness $0.1R_{\rm eq}$.  All values of $\Delta t$ are given relative to the mean outflow launch date of 2014 August 18.00 UT, inferred from the model. The uncertainties correspond to 1 standard deviation and are computed using a Markov Chain Monte Carlo approach.}
\end{longtable*}
\end{center}

\begin{figure}
\centerline{\includegraphics[width=3.5in]{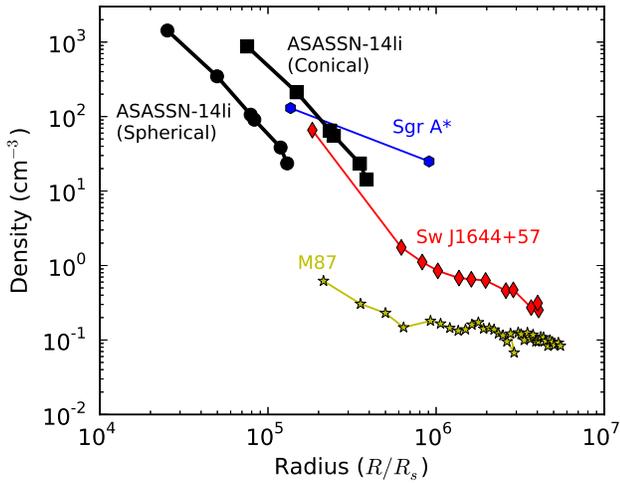}}
\caption{The radial density profile in the circumnuclear region of ASASSN-14li in comparison to other SMBHs. We infer a density profile of $\rho(R)\propto R^{-2.5}$ on a scale of about 0.01 pc.  For comparison, we show the density profiles for the Sgr A$^*$ \citep{bmm+03}, the nucleus of M87 \citep{rfm+15}, and the circumnuclear region of the $\gamma$-ray TDE Sw\,J1644+57 \citep{bzp+12}, which span the range of $\rho(R)\propto R^{-3/2}$ to $R^{-1}$.  To facilitate the comparison we scale the radii by the Schwarzschild radius of each SMBH ($R_s=2GM_{\rm BH}/c^2$, where $M_{\rm BH}$ is the black hole mass), using an estimate of $M_{\rm BH}\approx 10^6$ M$_\odot$ for ASASSN-14li \citep{hol15,mkm15}.  We find that for the circumnuclear region of ASASSN-14li the density profile is steeper than previously seen in the other SMBH systems, but the density normalization is comparable.}
\label{fig:n}
\end{figure}

We find that the kinetic energy of the outflow is $E_K\approx 4-10\times10^{47}$ erg and is constant in time, in agreement with the inferred free expansion of the ejecta, but distinct from the increasing energy as a function of a time observed in core-collapse SNe (c.f.~\citealt{bkc02}). Combining the outflow velocity and kinetic energy we infer an ejected mass of $M_{\rm ej}\approx 3\times10^{-5}-7\times10^{-4}$ M$_{\odot}$, dependent on the outflow geometry. This is $\sim 1-10\%$ of the mass accreted during the super-Eddington phase as inferred from modeling of the optical, UV, and X-ray emission (Figure \ref{fig:uvot}), consistent with theoretical estimates of the fraction of mass ejected in a wind during super-Eddington accretion \citep{sq09,lr11}.

We also find that independent of the outflow geometry, the pre-existing density profile in the circumnuclear region follows $\rho(R)\propto R^{-2.5}$ on a scale of $\sim 0.01$ pc (Figure~\ref{fig:n}), much smaller than the scale that can be directly probed in any extragalactic SMBH and even around Sgr A$^*$ \citep{bmm+03,rfm+15}.  The inferred profile is steeper than the $\rho(R)\propto R^{-3/2}$ profile expected for Bondi accretion in the circumnuclear regions of low accretion rate systems \citep{bon52}, and from the $\rho(R) \propto R^{-1}$ profile inferred within the Bondi radius of Sgr A$^*$ and the AGN in M87 \citep{bmm+03,rfm+15}.  The circumnuclear density profile inferred from radio observations of the relativistic TDE Sw\,J1644+57 is consistent with $\rho(R)\propto R^{-3/2}$ but shows a hint of a steeper slope at $R\simlt 0.05$ pc, the smallest radius probed \citep{bzp+12}.  The normalization of our inferred density profile depends on the outflow geometry, with $n\approx 60-500$ cm$^{-3}$ at a radius of 0.01 pc. This is comparable to the density found for Sgr A* and Sw\,J1644+57 at similar radii \citep{bmm+03,bzp+12}.

We note that the pre-TDE density inferred by our modeling is lower than the density required for spherical Bondi accretion at the rate implied by the archival observations \citep{bon52,vv15}. 
The calculated density increases somewhat if we assume that the system is not perfectly in equipartition (for example, if we use $\epsilon_B=0.01$ the overall density scale increases by about a factor of 5), but still falls short of the density required for Bondi accretion. However, this comparison relies on the assumption of spherical symmetry. In fact, simulations have shown that the density around an accreting black hole can be highly asymmetric, with densities in the plane of the accretion disk orders of magnitude higher than in the funnel carved out by a jet/outflow \citep{sn15}. It is likely that a jet existed prior to the onset of elevated accretion due to ASASSN-14li, as is typical of slowly accreting systems. If the outflow generated by the TDE was expelled along the same axis as the pre-existing jet, we could be probing this low-density funnel. Such an alignment is plausible if both outflows are aligned along the spin axis of the black hole. We therefore do not consider the inferred density to be problematic.  In fact, it may be indicative of alignment of the mildly collimated outflows before and after the TDE.

The model described above assumes that synchrotron and Compton cooling are unimportant.  With the parameters inferred from our radio observations for ASASSN-14li we expect these cooling breaks to be located at $\nu_c\simgt 10-20$ GHz, which is greater than $v_a$ and hence self-consistent with the model results. The precision of this calculation is limited by uncertainties in the the age of the outflow and propagated errors from uncertainties in the peak flux and peak frequency, but for any reasonable combination of parameters, the cooling breaks rapidly move to high frequencies during the span of our observations. Our January high-frequency flux deficit (see Figure \ref{fig:sed}) may be due to a cooling break, but may also be due to calibration errors arising from the fact that the VLA was in an intermediate configuration during that time, with larger uncertainties in the antenna position that will affect the high-frequency data. We also see a high-frequency flux deficit in our September observations, but this cannot be due to a cooling break because we see no evidence of such a break at lower frequencies in earlier epochs. There are no obvious calibration errors in the September high-frequency observations, so it is possible that the deficit may arise from some other mechanism. We note that this deficit does not affect our analysis, as the only quantities we need are the peak flux density and the frequency at which it occurs for each epoch.  Additional effects that reduce the high-frequency flux, while interesting, will not affect the main results of our analysis.

The synchrotron equipartition model readily generalizes to the case of relativistic expansion, with the bulk Lorentz factor of the outflow ($\Gamma$) as an additional parameter \citep{dnp13}.  In this case, to reach a self-consistent result in which $\Gamma\simgt 2$ (i.e., the outflow is relativistic) requires an unreasonably small value of $f_A$ that corresponds to a jet with an opening angle of $\simlt 0.1^{\circ}$.  This is two orders of magnitude narrower than the typical jets in GRBs \citep{fks+01}, and it would require fine-tuning in the jet orientation relative to our line of sight of $\sim 1.5\times 10^{-6}$ in order to detect the radio emission.  We therefore conclude that for any reasonable geometry the outflow from ASASSN-14li is non-relativistic.

\subsection{Interstellar Scintillation}

Using the inferred angular size of the outflow ($\theta_s\approx 8-80$ $\mu$as), we consider whether the observed radio emission might be affected by interstellar scintillation, which could lead to frequency- and time-dependent random variations in the radio flux density \citep{w98,gn06}.  Using the NE2001 Galactic free electron energy density model \citep{cl02}, we find that for the line of sight to ASASSN-14li the transition frequency between strong and weak scintillation is about 7 GHz, in the middle of our observation band.  At $\nu\simgt 7$ GHz we find that the fractional modulation level ($m_p$) due to ISS is at most a few percent (decreasing from $m_p\sim 10\%$ in our earliest 22.5 GHz observation to $m_p\sim 2\%$ in our final one).  However, at $\nu\simlt 7$ GHz we find an expected level of variation of up to $\sim 25\%$ at 1.45 GHz.  The 2015 August/September 1.45 GHz flux density presented in Figure 1 is an average of two observations obtained about 10 days apart.  Prior to averaging, the two epochs exhibit a $\sim 20\%$ flux density variation, consistent with the estimated effect of ISS.  This provides an independent confirmation of the small source size inferred from the equipartition analysis.

To verify that ISS-induced flux density variations do not bias our results, we repeated our equipartition analysis with larger errorbars on each data point, computed by adding in quadrature the measurement uncertainties and the expected ISS-induced modulation.  We find that while this increases the uncertainty on the derived physical properties of ASASSN-14li, the best-fit parameter values change by at most a few percent for the epochs with broad frequency coverage.

\subsection{Inconsistencies of a Single Component Model for the Radio Flux}\label{sec:1comp}

In Figure 2, we show the radial and time evolution of the model parameters derived from fitting the total radio flux (dotted lines) and the transient component only (solid lines).  The fits to the latter give a constant energy and velocity as a function of time, indicating that the outflow is in free expansion ($R_{\rm eq}\propto t$).  The outflow should continue expanding freely until it has swept up an amount of mass equal to its own initial mass. We can compute the amount of mass swept up from our derived density profile and we find that this is 
less than the inferred mass of the outflow, $M_{swept} \sim$ (0.04-0.4)$M_{ej}$ depending on the assumed outflow geometry. (In fact, $M_{swept}$ may be an even smaller fraction of the total outflow mass because we use the equipartition energy $E_{eq}$ to estimate $M_{ej}$, and $E_{eq}$ is the minimum energy of the system.) This result provides a self-consistency check for our model since the parameters are inferred from fitting the individual radio SEDs without an assumed temporal evolution.  Given the inferred steep density profile, we expect that the outflow will continue to expand freely for years to decades.

\begin{figure*}
\begin{center}
\includegraphics[width=6.4in]{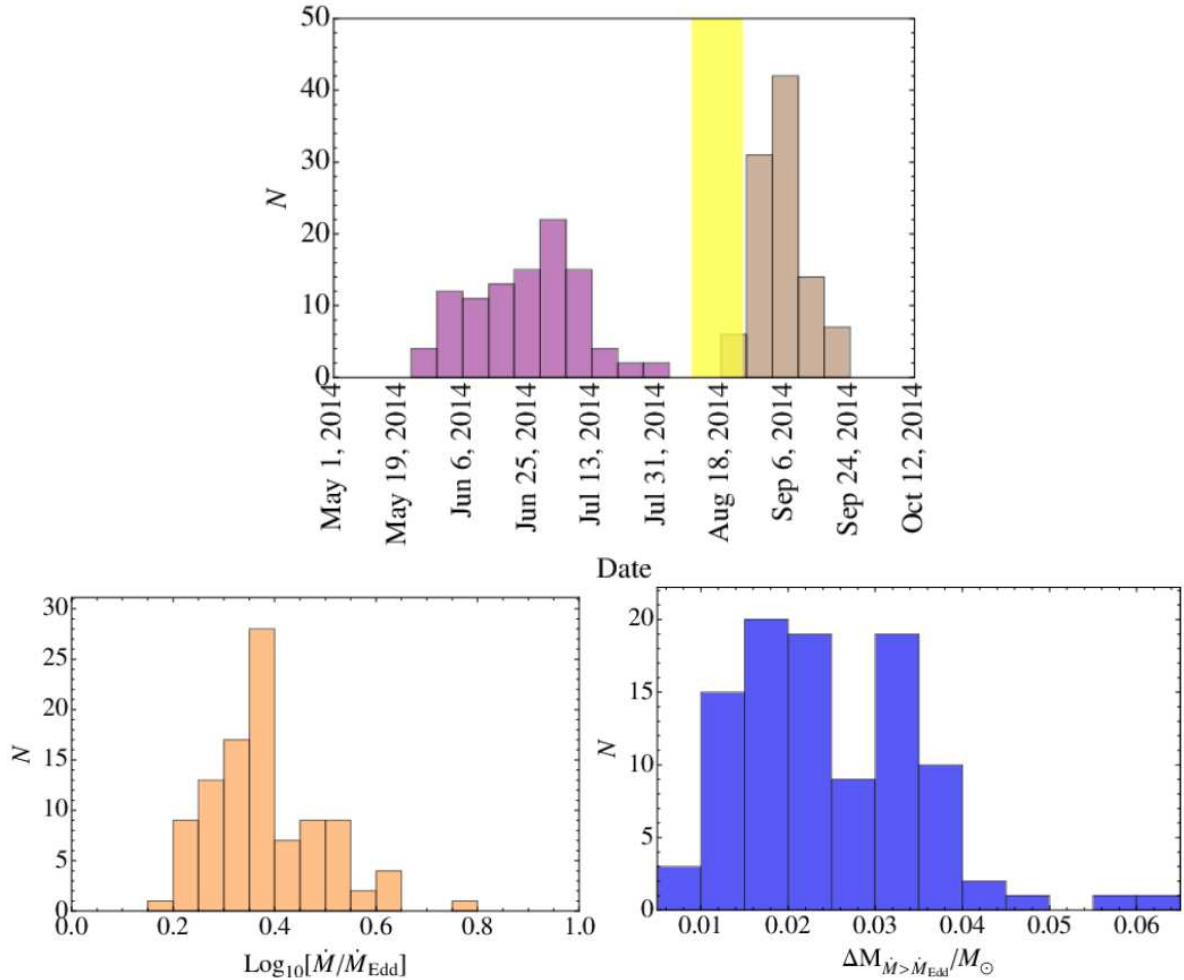}
\end{center}
\caption{Accretion parameters for ASASSN-14li estimated from modeling of the optical, UV, and X-ray observations.  (a) Histogram of the accretion milestone dates for the ensemble of model fits as compared to our determination of the outflow launch date (yellow band).  The purple histogram shows the time when each realization in the ensemble of model fits first crosses the Eddington limit, and the brown histogram shows the time when each realization reaches its maximum accretion rate. We find good agreement between our inferred outflow launch date and the times of super-Eddington and peak accretion. (b) Histogram of the maximum accretion rate normalized to the Eddington accretion rate ($\dot{M}_{\rm Edd}$) for each realization in our ensemble of model fits to the optical/UV light curves.  We find that ASASSN-14li exceeded the Eddington accretion rate by about a factor of 2.5. (c) Histogram of the total amount of mass accreted during the super-Eddington phase for each realization in our ensemble of model fits. The outflow mass that we infer from our radio observations is about $1-10\%$ of this total, in line with theoretical expectations.}
\label{fig:uvot}
\end{figure*}

In contrast, modeling of the total radio flux with a single component leads to energy and velocity evolution that are less natural.  The model fits imply that the outflow energy is increasing with time and that the outflow is accelerating, with $R_{\rm eq}\propto t^{1.6}$.  In core-collapse supernovae the kinetic energy is observed to increase with time due to the existence of ejecta at progressively slower velocities, with a steep profile of $E_K\propto v_{\rm ej}^{-5.2}$ \citep{tmm01}, but the velocity decreases with time.  The same is true for the behavior inferred from radio observations of the relativistic $\gamma$-ray TDE Sw\,J1644+57, in which an episode of energy increase by an order of magnitude was accompanied by a declining velocity \citep{bzp+12}.  Furthermore, an epoch-by-epoch comparison of the model fits to the total flux and to only the transient flux show that the total flux is not as well-fit by the synchrotron model, especially in our April 2015 observations (Figure \ref{fig:sed}).  For these reasons, and the archival radio detections, we conclude that the two-component model is correct, but we note that the overall main conclusion of a non-relativistic outflow is robust to our choice of model.

\section{Comparison with Other Modeling}
In this section we compare our results to independent modeling of the X-ray, UV, and optical observations (Guillochon et al. in prep) and consider alternate explanations for the radio emission. We find that our interpretation of the emission as a non-relativistic outflow launched during the period of super-Eddington accretion onto the SMBH is robust.

\subsection{Independent Modeling of the Accretion Rate from X-ray/UV/Optical Observations}\label{sec:uvot}

To determine the times at which the Eddington accretion limit is exceeded and when peak accretion is achieved, as well as the peak accretion rate and the total mass accreted in the super-Eddington phase we fit the optical, UV, and X-ray data of ASASSN-14li using the code {\tt TDEFit}; the data we fit against are the same data presented in \cite{mkm15} (see their Figure 1). Because the fallback of matter onto a black hole following a disruption only follows the canonical -5/3 law for half of disruptions, and only several months after the peak fallback rate \citep{gr13}, the fitting of tidal disruption light curves using a Monte Carlo approach is a far more robust procedure for constraining important temporal milestones for a given flare, such as the time of disruption and when the accretion rate crosses various thresholds such as the Eddington limit. {\tt TDEFit} utilizes a maximum-likelihood analysis to determine the most likely combination of disruption parameters, with one of the products being an ensemble of accretion rates onto the SMBH as functions of time. We find that the most likely black hole mass is $\approx 10^6 M_{\odot}$, and that the peak accretion rate is significantly in excess of the Eddington limit (Figure~\ref{fig:uvot}). 

Our modeling includes both the effects of inefficient circularization, which simulations have found significantly reduces the accretion rate onto the black hole relative to the fallback rate \citep{gmr14, skc15}, and limits the luminosity of the disk component to the Eddington limit. We find that the best-fitting circularization time is roughly three times longer than the timescale of peak accretion, resulting in a time of disruption that occurs much earlier than in models in which the viscous effects are neglected; this is the expected behavior for low-mass black holes ($M_{\rm BH} \sim 10^6 M_\odot$) where circularization takes place at large distances from the black hole \citep{gmc15}. This also reduces the peak accretion rate onto the black hole and imposes deviations from the canonical -5/3 decay law. We also find that the Eddington limit we impose reduces the luminosity of the flare significantly near the time of peak accretion onto the black hole, resulting in a reduced efficiency of conversion of accretion energy into observable optical/UV emission at these times. Our modeling is completely consistent with the early-time photometric limits for ASASSN-14li presented in \cite{hol15}.

Because our radio observations indicate that the outflow is in free expansion, we can extrapolate the observed radius to estimate $t_0$, the time at which the outflow was launched.  The launch time depends only weakly on the outflow geometry; we obtain $t_0=2014$ August $21$ ($\pm4$ days) for the spherical outflow ($f_A=1$) and $t_0=2014$ August $15$ ($\pm4$ days) for a conical outflow ($f_A=0.1$).  This time range is shown in comparison to the results from modeling of the optical, UV, and X-ray data in Figure~\ref{fig:uvot}.  We find that the outflow was launched at a time that straddles the onset of super-Eddington accretion and the time of peak accretion.  This supports our conclusion that the radio emission is due to an accretion-driven wind rather than being associated with the unbound debris, which would have been launched months earlier at the time of disruption. Figure~\ref{fig:uvot} also shows the total mass accreted during the super-Eddington phase as inferred from modeling of the optical, UV, and X-ray emission. Our estimate of the outflow mass is $\sim$a few percent of this number, consistent with theoretical estimates of the fraction of mass ejected in a wind during super-Eddington accretion \citep{sq09,lr11}.  We defer further description of the modeling work to a future paper (Guillochon et al. in prep).

\subsection{Radio Emission from the Unbound Debris}

After a TDE, approximately half of the stellar debris will be unbound from the black hole. The unbound debris around a non-spinning black hole will be very narrow in most cases as the stream is self-gravitating for low-beta encounters \citep{k94,gmr14,cn15}. When it is self-gravitating, its cross-section actually shrinks as it leaves the vicinity of the black hole, and likely only begins homologous expansion at a distance of $\sim10^{16}$ cm. At this distance, the stream covers a solid angle of $((r/r_t)^{1/4}r_{star}q^{-1/6}r)/(4r^2) \sim 10^{-5}$ steradians \citep{gmc15}. When the stream is not self-gravitating (which only occurs for deep, rare encounters, $\beta \simgt 3$), the maximum spread is given by the spread in velocity, estimated to be 0.2 steradians for a $10^6M_{\odot}$ black hole \citep{sq09}. The addition of spin will not dramatically alter these numbers; as described by \cite{k12} the maximum difference in the velocity spread will be about a factor of 2 (but often times can be reduced by a factor of 2). 

In our model, the physical size of the emitting region is well constrained by the equipartition argument. (The total energy of the system is a very strong function of radius, so this size estimate is robust even if the system is not perfectly in equipartition.) Therefore, if we assume that the radio emission covers only a small solid angle, we must conclude that the emission is emitted at a larger radius from the central black hole. This also naturally leads to a larger velocity of the emitting material, as the same fractional increase in the size of the emitting region requires covering a larger distance in the same amount of time. A self-gravitating debris stream covering a solid angle of $10^{-5}$ steradians at a radius of $10^{16}$ cm would produce a flux orders of magnitude too small to explain the observed radio emission. If we keep this solid angle and allow the emission to occur at a larger radius, the inferred velocity of the emitting material is $\Gamma\sim$ 2-3, which is much too fast to correspond to the unbound debris. 

For a non self-gravitating stream, the velocities are more reasonable; indeed, a solid angle of 0.2 steradians is not much more concentrated than the conical $f_A=0.1$ case we consider here. In this case, apart from the rarity of such high-beta encounters, an additional issue is matching the overall energies. The total energy we infer corresponds to a very small amount of material ($\sim2\times10^{-5}M_{\odot}$ for the 0.2 steradians case), while the total mass of the unbound material is orders of magnitude larger for the disruption of a solar mass or even 0.1 solar mass star. Even if we assume that only the fastest-moving tail of the distribution of unbound debris produces the radio emission, as recently suggested by \cite{kro16}, the emission expected in this case would still require a density tens to hundreds of times higher than the density we compute to match our observed fluxes. While the density we derive by assuming perfect equipartition is, like the energy, a lower limit, it is difficult to explain such a large discrepancy. Furthermore, at such high densities, the radio flux would be decreased by other effects, such as free-free absorption, and would not match the SEDs we observe. An additional issue is one of timing. As stated above, if we assume that the outflow has been moving at a constant velocity then we obtain a launch date that corresponds to the onset of super-Eddington accretion -- several months after the time of disruption. (Given that the current estimated radius of the emitting region is $\sim10^5R_s$, assuming that the emission was launched at a few $R_s$ instead of $R=0$ does not change this calculation.) It therefore seems unlikely that the radio emission could be generated by the unbound debris for any plausible geometry of the initial star-SMBH encounter.

\subsection{Comparison with a Decelerated Jet Model}
\begin{figure}
\centerline{\includegraphics[width=4in]{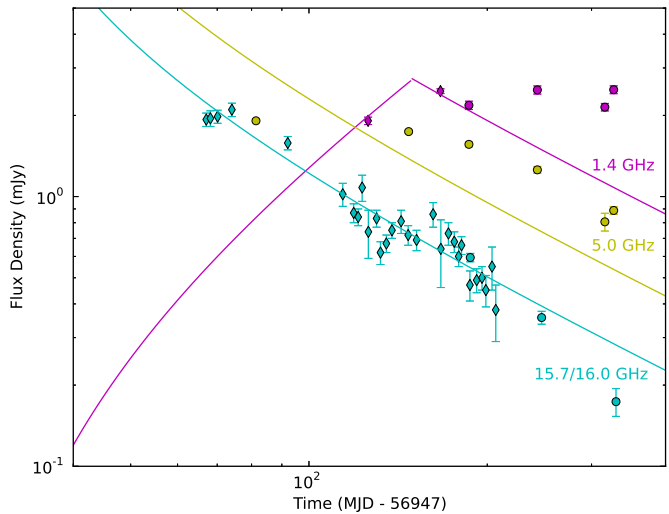}}
\caption{All currently available radio observations of ASASSN-14li at three representative frequency bands, as reported in \cite{vv15} (diamonds) and this work (circles). The solid lines show the expected flux evolution for the best-fit decelerated jet model presented in \cite{vv15}. (The time axis is chosen to match \citealt{vv15}'s Figure 2B.) We see that their model cannot reproduce our observed fluxes at 5.0 GHz and 1.4 GHz.}
\label{fig:vanvel}
\end{figure}

Our multi-frequency data rule out the interpretation of the radio emission as due to a decelerated (initially relativistic) jet, as recently proposed by \cite{vv15}. While their model provides a good fit to their observations, they are unable to constrain the evolution of $F_p$ and $\nu_p$ directly because most of their data is collected at a single frequency. This also means that they are forced to fix the circumnuclear density and density profile (which they assume to be flat). The density that they require to decelerate the jet at a radius of $10^{17}$ cm is much higher than the density we compute at that radius directly from our observations. In Figure \ref{fig:vanvel}, we present a modified version of their Figure 2B, which shows that their model does not fit our additional observations. Notably, their model predicts a steady decline in L band after ~March 2015, while we find that the total flux at 1.4 GHz remains roughly constant through September, with the exact level of variability difficult to quantify due to significant scintillation effects. The existence of a second steady-state component will not affect the quality of the model fit; subtracting the contribution of such a component would simply vertically shift all points at each frequency by the same amount.

\section{Conclusions}\label{sec:conc}

\begin{figure}
\centerline{\includegraphics[width=3.5in]{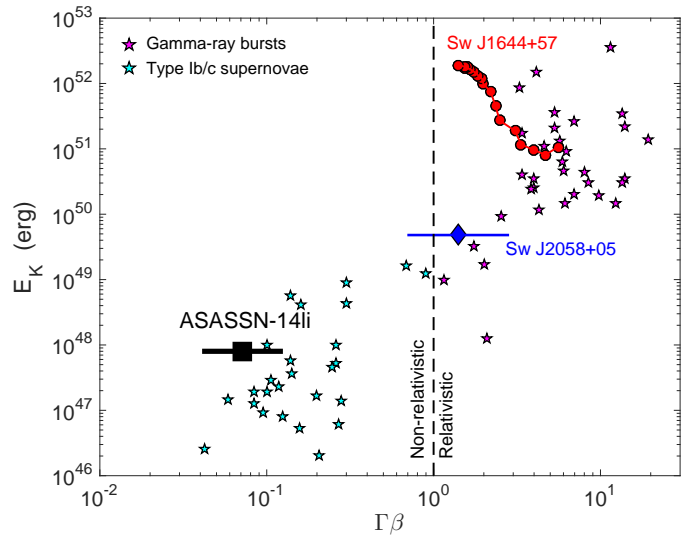}}
\caption{Kinetic energy ($E_K$) as a function of outflow velocity ($\Gamma\beta$) from radio observations of TDEs. We show the inferred values for ASASSN-14li (black square; horizontal bar represents the range of velocity for a range of outflow geometries) in comparison to the two $\gamma$-ray TDEs with radio emission: Sw\,J1644+57 (red circles; \citealt{zbs+11} and \citealt{bzp+12}) and Sw\,J2058+05 (blue diamonds; \citealt{ckh+12}).  The data for Sw\,J1644+57 are from detailed modeling of the radio emission as a function of time, including a correction for jet collimation with an opening angle of about 0.1 rad \citep{zbs+11,bzp+12}.  The data point and velocity range for Sw\,J2058+05 are based on an identical analysis to the one carried out here. The vertical dashed line at $\Gamma\beta=1$ roughly separates the phase-space into events with non-relativistic and relativistic expansion.  The $\gamma$-ray TDEs exhibit relativistic outflows with a large kinetic energy, but they represent $\simlt$ a few percent of the overall TDE volumetric rate \citep{mgm+15}.  On the other hand, ASASSN-14li exhibits a non-relativistic outflow with a lower kinetic energy but appears to represent the bulk of the TDE population.  Also shown for comparison are the data for long-duration $\gamma$-ray bursts (LGRBs; magenta stars) and Type Ib/c core-collapse supernovae (Type Ib/c SNe; cyan stars) \citep{mms+14}.  The LGRBs exhibit relativistic outflows with $E_K\simgt 10^{50}$ erg, while Type Ib/c SNe have non-relativistic outflows with $E_K\simlt 10^{49}$ erg.  In addition, LGRBs represent $\simlt 1\%$ of the Type Ib/c SN rate \citep{wp10}.  The TDE sample, although small, appears to trace the same relation seen in LGRBs and Type Ib/c SNe, with a small fraction of events (by volumetric rate) producing energetic relativistic outflows, and the bulk of the population producing lower energy non-relativistic outflows.}
\label{fig:context}
\end{figure}

We have detected transient radio emission associated with the nearby TDE ASASSN-14li, consistent with a non-relativistic outflow launched during the period of super-Eddington accretion. We conclude with several important implications of our results.  First, the velocity and kinetic energy of the outflow in ASASSN-14li are significantly lower than inferred for the two relativistic $\gamma$-ray TDEs previously detected in the radio (Figure \ref{fig:context}), which represent $\simlt$ a few percent of the TDE population \citep{zbs+11,bgm+11,bkg+11,mgm+15}.  Although the TDE sample with detected radio emission is small, this is reminiscent of the same relation observed in Type Ib/c core-collapse supernovae (Type Ib/c SNe) and long-duration gamma-ray bursts (LGRBs), in which a small fraction of events (LGRBs: $\sim 1\%$ by volumetric rate) produce energetic relativistic outflows while the bulk of the population (Type Ib/c SNe) produces lower energy non-relativistic outflows (Figure \ref{fig:context}; \citealt{mms+14}).

Second, ASASSN-14li is the nearest TDE discovered to date and the first to reveal radio emission associated with a non-relativistic outflow; previous upper limits on the radio luminosity of optical/UV TDEs are all at least a factor of a few above the level of emission detected here, and could only rule out the presence of relativistic jets \citep{bmc+13,vfk+13,cbg+14}.  This suggests that non-relativistic outflows are likely ubiquitous in most TDEs.  This conclusion is further supported by observations of the optical TDE PS1-11af at $z=0.405$ which revealed a broad rest-frame UV absorption feature with $v\sim 13,000$ km s$^{-1}$ suggestive of a similar outflow \citep{cbg+14}; such absorption was not detectable in other TDEs due to their lower redshift and hence lack of rest-frame UV spectral coverage.

Finally, given the likely ubiquity of outflows from most TDEs we expect such events to be detected in future sensitive wide-field radio surveys of the local universe; for example, the Square Kilometer Array will be able to probe a volume $\sim$100 times larger than that accessible to current facilities for a radio luminosity comparable to that of ASASSN-14li \citep{cr+04}. Time-series rest-frame UV spectroscopy of more distant TDEs may also serve to infer the presence of outflows and the timing of their ejection.

\begin{acknowledgements} K.D.A., E.B., and P.K.G.W.~are supported in part by NSF and NASA grants. J.~G.~ acknowledges support from Einstein grant PF3-140108. A.~Z.~ acknowledges support from NSF grant AST-1302954. The VLA is operated by the National Radio Astronomy Observatory, a facility of the National Science Foundation operated under cooperative agreement by Associated Universities, Inc.  \end{acknowledgements}


\end{document}